\documentclass[aps,prd,showkeys,superscriptaddress,nofootinbib,floatfix]{revtex4-2}
\usepackage[utf8]{inputenc}
\usepackage{hyperref}
\usepackage{amssymb,amsmath,amsfonts}
\usepackage{xcolor}
\usepackage{graphicx}
\usepackage{datetime}
\usepackage{adforn}
\usepackage{fourier}
\usepackage{bbding}
\usepackage{pifont}
\usepackage{soul}
\begin{document}
\title{Extracting more information from entropy}
\author{L. Araque}
\email{aralameda@gmail.com}
\address{Departamento de Física, Escuela de Ciencias, Núcleo de Sucre, Universidad de Oriente, Cumaná, Estado Sucre, Venezuela}
\author{W. Barreto}
\email{willians.barreto@ufabc.edu.br}
\address{Centro de Ciências Naturais e Humanas, Universidade Federal do ABC, Av. dos Estados 5001, CEP 09210-580, Santo André, São Paulo, Brazil}
\address{Centro de Física Fundamental, Facultad de Ciencias, Universidad de Los Andes, Mérida, Estado Mérida, Venezuela}
\begin{abstract}
We extract the complex frequency of the lowest quasi-normal mode (QNM) from the holographically computed entropy density near thermodynamic equilibrium. 
The system under study is a purely thermal Supersymmetric Yang-Mills $\mathcal{N}=4$ plasma undergoing homogeneous isotropization dynamics. 
Starting from a far-from-equilibrium initial state, the system evolves toward equilibrium entropy, forming a stairway pattern.  
Our analysis reveals that the rate of increase of the stairway is twice the decay rate (imaginary part) of the lowest QNM. Based on this observation, we propose a model that explains how this information is encoded in the entropy.
The model is further extended to include finite temperature, R-charge density, and scalar condensate, revealing an additional feature: the system's dominant dissipation channel may shift to one driven by the scalar condensate, depending on the chemical potential.
\end{abstract}
\maketitle  
\section{Introduction}
\textcolor{black}{Recent evelopments in Quantum Gravity and Quantum Information aim to unveil and unify the concept of entropy, simi\-lar to progress made with quantum entanglement.}
In that sense, it is instrumental the black hole thermodynamics \cite{w24} via holographic gauge-gravity duality \cite{m98}-\cite{w98b}, where entropy is well-defined and extensively studied. Holographic duality came up to connect Fluid Dynamics, High Energy Physics, Nuclear Physics, General Relativity and Condensed Matter, among other areas \cite{n14}.

In a recent study \cite{rb24} reported a stairway to equilibrium entropy for the 1RCBH model, which features a critical point in its conformal phase diagram at finite temperature and R-charge density. The evolution of the non-equilibrium entropy in a homogeneous isotropization dynamics, as well as the pressure anisotropy and the scalar condensate of the medium were computed. 
\textcolor{black}{For all analyzed initial data, there forms a periodic sequence of several close plateaus near thermodynamic equilibrium, creating a stairway pattern in the entropy density.}
The stairway step in time (tread and raiser) has a period which is a half the period of oscillations of the lowest quasi-normal mode (QNM) of the system. 
However, that study did not address the rate of entropy increase in the stairway. 
For the parti\-cular case of purely thermal Supersymmetric Yang-Mills (SYM) $\mathcal{N}=4$ plasma at zero R-charge density and vanishing scalar condensate, was found that the period of the stairway is half the period of oscillations of the lowest QNM associate to the late time equilibration of the pressure anisotropy of the fluid, while at finite chemical potential the lowest QNM of the system is associated to the late time equilibration of the scalar condensate.  

After a careful revision, \textcolor{black}{in this brief note we report additional features for 
the well studied homogeneous isotropization dynamics \cite{cy09}, \cite{wilke}, \cite{crn17}, \cite{rb24}}. 
\textcolor{black}{We go \textcolor{black}{further in} trying to understand an \textcolor{black}{unexpectedly} complex and \textcolor{black}{subtle} system}. In order to study more deeply that system and gain insight, here we start with the stairway to the equilibrium entropy (for the purely thermal SYM $\mathcal{N}=4$ plasma), with the numerical solver output data of \cite{rb24}. For that reason, the equations (and algorithms to solve them) are not presented here. Close to equilibrium, the structured data is pe\-riodic and increasing in time for any initial data that was considered and evolved. 
We extract the information which characterizes the stairway, resulting equal to twice the real and imaginary parts of the well known complex frequency of the lowest quasi-normal mode for the pressure anisotropy. \textcolor{black}{This leads us to propose a model that allows obtaining the stairway from the pressure anisotropy.} 
We extend the analysis to consider finite temperature, R-charge density and scalar condensate of the medium, besides pressure anisotropy. We find that the channel of dissipation may change in time, depending on the chemical potential. 
We also use here the natural units where $c=\hbar=k_{B}=1$.

\section{Extraction}\label{sec:2}
\begin{figure}
\includegraphics[width=0.4\textwidth]{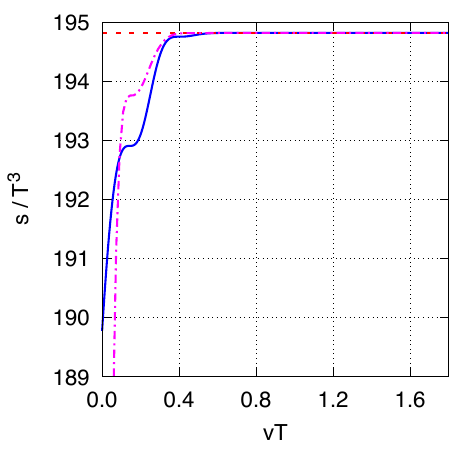}
\caption{Evolution of dimensionless non-equilibrium entropy density (blue line) as computed in Ref. \cite{rb24} using initial data (IC 4) given by Eq. (37) and parameters specified in Table I (with $Q=0$). Observe that there are not apparent changes beyond ${vT}\approx 0.6$. The red dashed line corresponds to the equilibrium dimensionless entropy density. \textcolor{black}{The magenta dot-dashed line corresponds to the entropy density as modeled in Section III}.}
\label{fig:figure1}
\end{figure}

\begin{figure}
\includegraphics[width=0.4\textwidth]{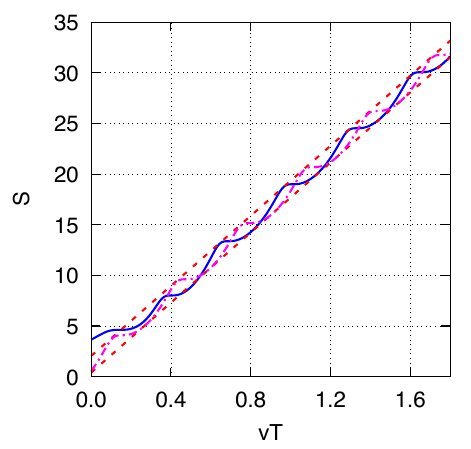}
\caption{$S$ (blue line) as a function of time (the stairway), for the same initial condition and parameters informed in Fig. \ref{fig:figure1}. Observe that before ${vT}\approx 0.6$ the fluid is far-from-equilibrium, possibly in a non-linear regime. 
\textcolor{black}{The referential red dashed lines have slopes twice 8.64 and are shifted to be tangents to the stairway by above and below; the frequency of the stairway is twice 9.81. The pair of numerical values $(9.81,8.64)/\pi\approx(3.12,2.75)$  corresponds to the well known lowest QNM for the purely SYM case \cite{s02}.
The magenta dot-dashed line is the stairway as modeled in Section III}.} 
\label{fig:figure2}
\end{figure}

\begin{figure}
\includegraphics[width=0.4\textwidth]{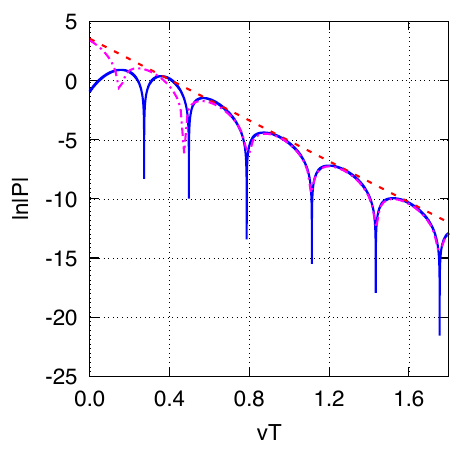}
\caption{$\ln|P|$ (blue line) as a function of time, for the same initial condition and parameters informed in Fig. \ref{fig:figure1}. The pressure anisotropy in its way to equilibrium behaves like a QNM with complex frequency $9.81 - i8.64$. The red dashed line is referencial and has slope $-8.64$. \textcolor{black}{The magenta dot-dashed line is $P$ given by Eq. (\ref{fit}) fitted to the holographically computed data to give $A=30.0$ and $B=1.65$. Observe that based on extraction (Section I) we have used $a=8.64$ and $b=9.81$}.}
\label{fig:figure3}
\end{figure}

We begin with the entropy density, $s$, computed  holographically in \cite{rb24} (with slight changes in  notation) and shown in figure \ref{fig:figure1}. For illustration, we use one initial condition.    
Defining 
\begin{equation}
S\equiv\ln\left\{\frac{s_{\text{eq}}}{s_{\text{eq}}- s}\right\}, \label{S}
\end{equation} 
where $s_{\text{eq}}$ is the equilibrium entropy density and $s$ is the Bekenstein-Hawking entropy density calculated for the apparent horizon, we plot in figure \ref{fig:figure2} the stairway to the entropy density  equilibrium, $S$. The periodicity and rate of increase of the stairway are clearly shown. \textcolor{black}{To extract the stairway period and rate of increase, we performed a numerical analysis within the corresponding time window. Specifically, we calculated the finite difference to identify the minima and maxima of $S$. Next, we utilized the well-known software {\sc harminv} \cite{harminv} to compute the mean period of these extrema by evaluating the second finite difference of $S$. Using the values of $S$ at the times corresponding to maxima (or minima) of the first difference, we applied the standard least-squares method to determine the rate of increase of the stairway, which was found to be $\approx 17.3$}. The referen\-tial dashed lines in \textcolor{black}{figure \ref{fig:figure2}} have slopes that are twice $8.64$ and are shifted to be tangents to the stairway by above and below. As was established in \cite{rb24} the period of the stairway for this specific case is $\approx 0.32$ which corresponds to twice the angular frequency $9.81$. The numerical pair $(9.81,8.64)/\pi\approx(3.12,2.75)$  corresponds to the well-known lowest QNM for the purely SYM case \cite{s02}. Now, we define  
\begin{equation}
P\equiv\frac{\Delta p}{\epsilon},\label{anip}
\end{equation}
where $\Delta p$ is the pressure anisotropy and $\epsilon$ the energy density \cite{rb24}.
In the interval of extraction the stairway plateau is not in phase with zero pressure anisotropy, as displayed in figure \ref{fig:figure3}.
\begin{figure}
\includegraphics[width=0.4\textwidth]{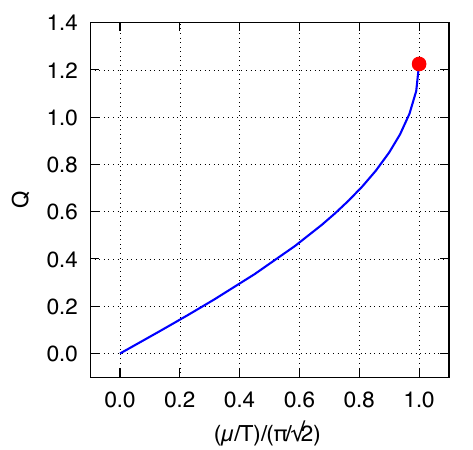}
\caption{Black hole charge $Q$ as a function of the normalized $\mu/T$. The red point represents the critical point at $\mu/T=\pi/\sqrt 2$.}
\label{fig:figure4}
\end{figure}
\begin{figure}
\includegraphics[width=0.4\textwidth]{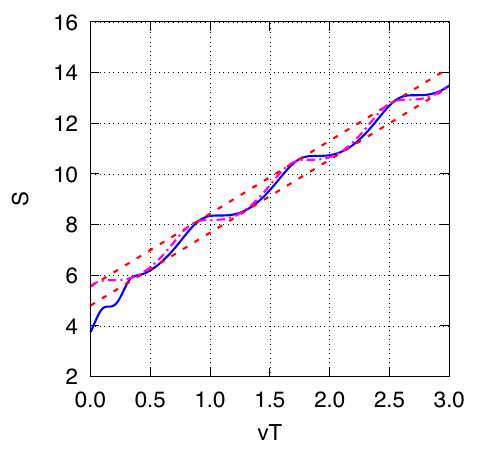}
\caption{$S$ (blue line) as a function of time (the stairway), for the same initial condition and parameters informed in Fig. \ref{fig:figure1}, but now for $\mu/T=\pi/\sqrt{2}$. \textcolor{black}{The magenta dot-dashed line is the modeled stairway}, but now using Eq. (\ref{epro_}). For this plot the fitting parameters are $A=4.95$, $B\approx-0.70$, $\mathcal{K}\approx0.30$, considering the extracted values of the lowest complex frequency for the singlet channel
\footnote{\textcolor{black}{DeWolfe, Gubser and Rosen \cite{dgr11} identified diffeormorphism and gauge-invariant linear combinations of the Einstein-Maxwell-Dilaton (EMD) field fluctuations which are classified into three different irreducible representations of the $SO(3)$ rotation group: the quintuplet (spin 2), the triplet (spin 1), and the singlet (spin 0) channels.}}
$3.81-i1.44$ \cite{crn17}. \textcolor{black}{The referential red dashed lines have slopes twice $1.44$ and are shifted to be tangents to the stairway by above and below; the frequency of the stairway is twice $3.81$.}} 
\label{fig:figure5}
\end{figure}
\begin{figure}
\includegraphics[width=0.4\textwidth]{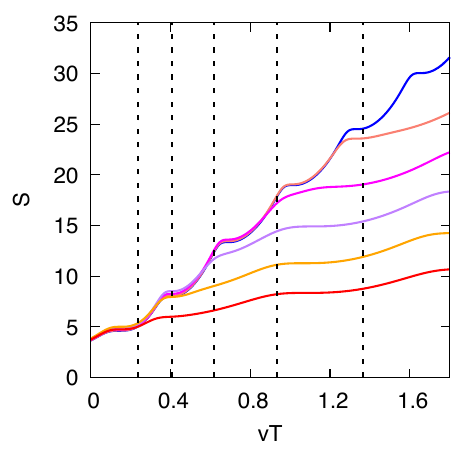}
\caption{The stairway $S$ as a function of time, for the same initial condition and parameters informed in Fig. \ref{fig:figure1}, but now for different values of $\mu/T$: 0.00 (blue); 0.50 (salmon); 1.00 (magenta); 1.50 (purple); 2.00 (orange); including the value for the critical point, $\mu/T=\pi/\sqrt 2\approx 2.22$ (red). The vertical dashed lines indicate the \textcolor{black}{estimated crossing} time in which $\ln|\Phi|>\ln|P|$.}
\label{fig:figure6}
\end{figure}

\section{Explanation}\label{sec:3}
Based on the extraction, we propose the following ansatz 
\begin{equation}
\frac{d\hat s}{d\tau}=\mathcal{K}P^2, \label{epro}
\end{equation}
where $\hat s=s/T^3$, $\tau=vT$ and $\mathcal{K}$ a constant to be determined. Fitting $P$, given by
\begin{equation}
P=Ae^{-a\tau}\sin(b\tau+B), \label{fit}
\end{equation}
to the computed holographically data, we get $A=30.0$ and $B=1.65$. Note that, based on the previous section, we have used $a=8.64$ and $b=9.81$. The magenta curve in figure \ref{fig:figure3} represents this fit. Integrating (\ref{epro}) we obtain
\begin{equation}
\hat s=\hat s_0+\frac{K(\tau)}{4} \label{emod},
\end{equation}
where the area is given by
\begin{equation}
K(\tau)=\frac{\mathcal{K}A^2}{(a^2+b^2)}e^{-2a\tau}\left\{a\cos[2(b\tau+B)]-b\sin[2(b\tau+B)]-\frac{a^2+b^2}{a}\right\},
\end{equation} 
which can be easily written as 
\begin{equation}
K(\tau)=\frac{\mathcal{K}A^2}{\omega_\ell}e^{-2a\tau}\left\{\cos[2(b\tau+B)+\delta]-\frac{\omega_\ell}{a}\right\},\label{area}
\end{equation}
where $\omega_\ell=(a^2+b^2)^{1/2}$ and $\delta=\arctan(b/a)$. Cosenquently, Eq. (\ref{epro}) complies with the second law of thermodynamics and agrees with the findings in \cite{jm16}. In particular, Eq. (\ref{emod}) coincides with Eq. (1) of \cite{jm20}, which addresses a more general context.
Using the correspondent equilibrium value for $\hat s_0$ we get $\mathcal{K}\approx\pi$. The magenta curve in figures \ref{fig:figure1} and \ref{fig:figure2} represents the modeled entropy density and stairway, respectively. Observe that the modeled stairway is not in phase with the holographic computed curve (in blue). To be in phase is required an additive phase of $\approx 1.2$. Thus, our model explain the stairway structure but not its phase deficit/excess, possibly of non-linear origin. The linear response can be obtained also from a perturbative analysis \cite{note}, but here we got it from data analysis. 

\section{Extension}
Now we consider the model 1RCBH \cite{rb24}. To maintain simplicity, we define \textcolor{black}{the scalar condensate anisotropy as}
\begin{equation}
\Phi\equiv\frac{\Delta\phi}{T^2}, \label{epro_}
\end{equation}
where $\Delta\phi=\langle O_\phi\rangle-\langle O_\phi\rangle_{\text{eq}}$, being $\langle O_\phi\rangle$ the scalar condensate; the extra subscript stands for equilibrium.  
In this model we have a phase diagram with a critical point, as illustrated in figure \ref{fig:figure4}. $Q(\mu/T)=0$ is the particular case considered in sections \ref{sec:2} and \ref{sec:3}, that is, purely SYM with zero R-charge density and zero scalar condensate. For the same initial condition we consider now other chemical potentials $\mu/T$, including the value for the critical point. Figure \ref{fig:figure5} shows the stairway computed holographically for the critical point, $\mu/T=\pi/\sqrt{2}$, and the stairway modeled using the fitting parameters for $\Phi$. Observe that the phase shift between stairways is less and negative in comparison with the phase shift for the pressure anisotropy (see figure \ref{fig:figure2}). Figure \ref{fig:figure6} shows transient features not studied in \cite{rb24}, that is, the change of the increasing rate of the stairway for the same initial condition and different values of chemical potential. The model implemented in section \ref{sec:3} can be applied to the scalar condensate anisotropy given by Eq. (\ref{epro_}). But why this change in the stairway design? It can be inferred from each stairway that the time in which the rate of increase changes is marked by the time in which the pressure anisotropy cross (to be less than) the scalar condensate anisotropy. A practical implementation of this is considering $\ln|\Phi|>\ln|P|$. We confirm that by means of the estimated  crossing times (from the anisotropies data computed holographically) represented by the vertical lines in figure \ref{fig:figure6}.       
It is interesting to note that even far-from-equilibrium this picture works.

\section{Conclusions}
In this work we analyze in greater detail the stairway to equilibrium entropy for the homogeneous isotropization dynamics of the top-down 1RCBH holographic model, as computed in \cite{rb24}. 

From the entropy density, we extracted the lowest QNMs of the system, which correspond to the well-known complex frequencies in the literature. Notably, the stairway encodes twice the lowest QNM associated with the singlet and quintuplet channels \cite{crn17}, depending on the chemical potential and the evolution of time.  To explain the stairway structure for any chemical potential, including the critical point of the phase diagram, we propose a model. 
\textcolor{black}{While the model successfully explains the stairway structure, it does not account the phase deficit/excess observed in the stairway. However, using holographic duality, we estimated that deficit/excess of the phase shift,} \textcolor{black}{ which lies beyond the linear response of the system.} 
\begin{figure}
\includegraphics[width=0.4\textwidth]{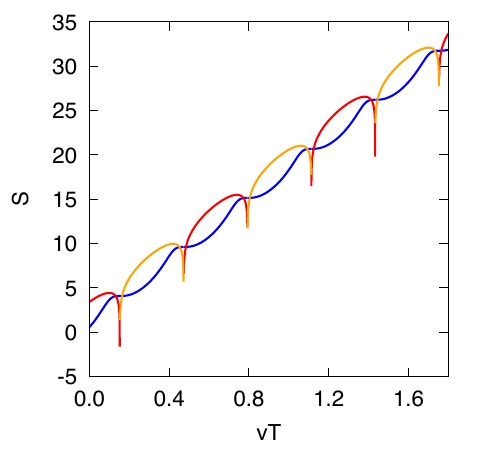}
\caption{Modeled stairway $S$ (blue curve) as a function of time, for the same initial condition and parameters informed in Fig. \ref{fig:figure1}. For this plot the fitting parameters are $A=30.0$, $B=1.65$, $\mathcal{K}=\pi$, considering the extracted values
of the lowest complex frequency of the quintuplet channel $b-ia=9.81 - i8.64$ \cite{crn17}. The superposed segments correspond to $2a\tau +\ln|A\sin(b\tau +B)|$: $P>0$ (red) and $P<0$ (orange). Each plateau of the stairway  begins to form when $|P|$ {\it isotropizes} and begins to rise when $|P|$ {\it anisotropizes}.} 
\label{fig:figure7}
\end{figure} 
\textcolor{black}{Our findings demonstrate that the system's dominant dissipation channel may shift to one driven by the scalar condensate, depending on the chemical potential.}

\textcolor{black}{The proposed model provides an explanation  of the plateaus formation and the rate of entropy increase}. If isotropization is understood as the tendency of anisotropy to approach zero, each plateau forms when the system transiently isotropizes. In this state, entropy production approaches zero as dissipation (anisotropy) diminishes. In turn, entropy production reaches its maximum when the anisotropy reaches an extremum. This behavior is illustrated in figure \ref{fig:figure7}. A similar mechanism applies to the stairway associated with the scalar condensate. From the gravitational dual perspective, the formation of a plateau corresponds to the absence of gravitational or scalar radiation flux to the black hole. \textcolor{black}{In this sense, Eq. (\ref{epro}) \textcolor{black}{resembles the} power radiated \textcolor{black}{at time $\tau$ across the balck hole's horizon area} \cite{b14}.} 
When the system reaches maximum (or minimum) anisotropy, the entropy production rate peaks. Thus, isotropy is isentropic, while anisotropy is dissipative \cite{hdo24}. 
Far-from-equilibrium, the mechanism driving isotropization  appears to resemble a QNM channel. The non-linear evolution regime, however, requires further investigation to fully explain the symmetry breaking in plateaus and phase shift with relative to the QNM zeros.

\textcolor{black}{Several key points are worth emphasizing. We analyze a non-linear data for the homogeneous isotropization of a holographic 1RCBH model, focusing on the  the late-time evolution of the entropy density. The stairway is modeled using the ansatz in Eq. (\ref{epro}). Our findings  clearly demonstrate how the lowest complex frequency of the QNM is encoded in the stairway. However, the first iteration of the model does not explain a significant deficit/excess in phase shift, which must be determined numerically. This deficit/excess in phase shift occurs far-from-equilibrium, where the system is in the non-linear regime. Interesting,  phase shift remains constant once the system enters the linear regime, closer to equilibrium. Our model is based on non-linear data analysis rather than perturbative methods. It suggests that any fluid undergoing homogeneous isotropization, starting far-from-equilibrium,  will approach  equilibrium by forming an entropy stairway that encodes the lowest complex QNM of the dominant dissipative channel. This hypothesis should be tested with other models, which is beyond of this brief report.}

\textcolor{black}{Close to equilibrium, the collective (macroscopic) response of the plasma manifests as the lowest QNM in the pressure anisotropy and the scalar condensate anisotropy, defined by Eqs. (\ref{anip}) and (\ref{epro_}), respectively. This response is mirrored in the plasma entropy density. It is worth noting that the anisotropies (of pressure and scalar condensate) and the entropy density considered in our study were obtained holographically for the 1RCBH model in [7], mapping the strongly coupled plasma dynamics to a gravitational system (containing a black hole) described by the EMD equations. On the other hand, microscopic studies of QNMs and entropy, and their  correlation for a Quark-Gluon Plasma (QGP) and its proxies, are well-established in the literature (see \cite{lh25} and references therein). Building on this foundation, our work provides a quantitative realization of the connection between QNMs and entropy production. These findings represent a significant step forward in understanding the role of QNMs in holographic homogeneous isotropization dynamics.}
 
The proposed model extends beyond linearity. Specifically, the deficit/excess phase shift observed in the entropy stairway, which cannot be explained by linear response theory, suggests the presence of nonlinear effects. This study reveals that dissipation mechanisms are governed by QNMs, represented by the gravitational and scalar radiation flux to the black hole's interior in the bulk. On the dual boundary, these mechanisms are driven by the pressure anisotropy or scalar condensate anisotropy, depending on the chemical potential $\mu$. While our study builds on existing numerical data, it provides new insights into the connection between QNMs and entropy production, offering a framework that can be extended to other holographic models or non-linear regimes. This focused case study lays the groundwork for future investigations into more complex systems, provided they describe homogeneous isotropization processes. However, our analysis is restricted to the 1RCBH model dynamics. 

On physical grounds, it is reasonable to expect that any homogeneous isotropization model would exhibit a stairway to equilibrium with an encoded fundamental QNM.
\textcolor{black}{Finally, we note that the extraction of non-hydrodynamic QNMs for the Bjorken flow can be achieved by subtracting attractors \cite{gb24}. In that case, a tail dominates the late time evolution, characterized by a purely imaginary frequency. For other homogeneous isotropization models, it is interesting to consider one with purely imaginary frequencies in its spectra \cite{dr24}, \cite{dbr25}.}
%\footnote{\textcolor{black}{We know about an ongoing work on the 2RCBH model which will help test the generality of our findings.}}.

\textcolor{black}{We highlight the need to explore non-linear effects beyond the deficit/excess phase shift far-from-equilibrium, which are not captured by the model presented here. In this sense could useful to consider non-linear models of higher order than (\ref{epro}) and methods as Deep Neural Networks and Tensor Networks.}
In this work, we extracted additional information from entropy. As stated in \cite{rb24}, it seems to be possible, in principle, to predict general features \textcolor{black}{of} one observable (entropy/anisotropy) by knowing the other (anisotropy/entropy). 
\textcolor{black}{We believe this kind of analysis could be extended to consider other models in future studies.}
{\acknowledgements}
\textcolor{black}{We thank Nairy Villarreal, Rômulo Rougemont, Carlos Peralta and Beltrán Rodríguez for reading the first version of the manuscript. WB thanks FAPESP, Research Projects Program, under grant 2022/02503-9 and acknowledge the financial support by National Council for Scientific and Technological Development (CNPq) under grant number 407162/2023-2. Also thanks to the {\it Central de Computação Multiusuário} (CCM) at UFABC, for support using the clusters {\it Titânio} and {\it Carbono}.}

\end{document}